\documentclass[10pt,amsmath,aps,prb,superscriptaddress,showpacs,amssymb,secnumarabic,eqsecnum,twocolumn,nofootinbib,floatfix]{revtex4-1}
\usepackage[dvips]{graphicx,color}
\usepackage{graphics}
\usepackage{pstricks}
\usepackage{epstopdf}
\usepackage{graphicx}
\usepackage{amssymb}
\def\ba{\begin{eqnarray}}
\def\ea{\end{eqnarray}}
\def\be{\begin{equation}}
\def\ee{\end{equation}}
\def\n{\nonumber\\}

\begin{document}
\title{Approximate analytic expression for the Skyrmions crystal} 
\author{Nicol\'as Grandi}
\affiliation{Instituto de Fisica de La Plata - CONICET \& Departamento de Fisica - UNLP,\\ C.C. 67, 1900 La Plata, Argentina}
\author{Mauricio Sturla}
\affiliation{Instituto de Fisica de La Plata - CONICET \& Departamento de Fisica - UNLP,\\ C.C. 67, 1900 La Plata, Argentina}
\begin{abstract}
We find approximate solutions for the two-dimensional non-linear {$\Sigma$}-model with Dzyalioshinkii-Moriya term, representing magnetic Skyrmions. They are built in an analytic form, by pasting different approximate solutions found in different regions of space. We verify that our construction reproduces the phenomenology known from numerical solutions and Monte Carlo simulations, giving rise to a Skyrmion lattice at an intermediate range of magnetic field, flanked by spiral and ferromagnetic phases for low and high magnetic field respectively.
\end{abstract}
\maketitle
\section{Introduction}
The emergence of the magnetic Skyrmion structure in condensed matter is well known
since long time ago. They appear in a variety of systems such as
liquid crystals \cite{WM_89}, quantum-Hall ferromagnets \cite{SKK_93} and Bose
condensates \cite{Ho_98}.

These topological nano-sized spin textures have a long lifetime, and
had been observed by Lorentz transmission electron microscopy
\cite{YOK_10}, by neutron scattering \cite{MBJ_09}, and by
spin-resolved scanning tunneling microscopy
\cite{HvBM_11,Pleiderer_11}. Interesting properties related to
magnetic Skyrmions have been reported. Among others, the possibility
of driving the motion of the Skyrmions with ultra-low current density
\cite{JMP_10, NagaosaTokura2013}, becomes particularly interesting
because of their possible application in information storage and
processing devices \cite{SCR_13,TMZ_14}. The responsible for these
properties, seems to be the emergent electromagnetism associated with
the non-collinear spin structure of the Skyrmions
\cite{NagaosaTokura2013}.

Skyrmions are topologically non-trivial field configurations
characterized by an integer number $N_{sk}$, called Skyrmion number,
and they can be stabilized by different mechanisms. In anisotropic
easy-axis magnetic thin films, for instance, the presence of a
long-range dipolar interaction leads to a striped phase, that can
evolve to a periodic arrange of Skyrmions when a magnetic field is
applied perpendicular to the film \cite{LGG_73}. The size of these
Skyrmions range from $10$nm to $1\mu$m, and the lowest energetic
configuration is characterized by $N_{sk}=1$ and a helicity
$\gamma=\pm \pi/2$. On the other hand, Skyrmions can also appear in
presence of four-spin exchange interaction or when exchange
interaction is frustrated. In these cases, the characteristic scale is
of order of $1$nm, and Skyrmions, $N_{sk}=1$, and anti-Skyrmions,
$N_{sk}=-1$, are energetically equally favorable
\cite{OCK_12,HvBM_11}. Finally in non-centro-symmetric magnets a
Dzyaloshinskii-Moriya interaction \cite{Dzyaloshinsky_58,Moriya_60}
can stabilize a Skyrmion crystal phase, in a given window of
temperature and magnetic field. In this case, the typical size range
from $5$ to $100$nm, and they have $N_{sk}=1$, and helicity
$\gamma=\pm \pi/2$. The sign of $\gamma$ is fixed by the sing of $D$,
the Dzyaloshinskii-Moriya interaction strength, which is determined by
the crystal structure. This mechanism seems to be the responsible for
the observation of Skyrmion crystal structures in chiral magnets, such
as those reported in, FeGe\,\cite{LBF_89,UNH_08,YKO_11,WBS_11}, MnSi\,
\cite{MBJ_09,ITB_76,IA_84,GDM_09,LHP_95,PRP_04,JGM_13},
Mn$_{1-x}$Fe$_x$Ge\, \cite{SYH_13}, and
Fe$_{1-x}$Co$_x$Si\,\cite{BVR_83,GDM_07,GCD_09,OTT_05}.
These chiral magnets seem to be the natural arena to develop
technological applications of Skyrmions, particularly to
racetrack-type memories for logic computing technologies
\cite{SCR_13,TMZ_14}.

The existence of these topologically protected particles that can be
stabilized in chiral magnets was early theoretically predicted
\cite{BY_89,BH_94,RBP_06}. Later on, in a seminal work, Yi {\em
  et.al.} \cite{YON_09} showed by Monte Carlo simulation that a
$2$-dimensional classical spin system with a Dzyaloshinskii-Moriya
term and magnetic anisotropy, has a rich helical phase
diagram. Additionally they found that for a particular point of
parameter space, a Skyrmion lattice ground state emerge.

The low energy dynamics of classical spin system is given by a
non-linear sigma model. A well known topologically non-trivial
solution of the non-linear sigma model is the Skyrmion. In the
particular case of the square lattice with ferromagnetic exchange
interaction, the low energy dynamics is governed by the so-called
$O(3)$ non-linear sigma model. The order parameter space of the $O(3)$
non-linear sigma model is, actually, $O(3)/O(2)\simeq S^2$ that has a
non-trivial second homotopy group $\pi_2(O(3)/O(2))={\mathbb Z}$. Because
$\pi_2(O(3)/O(2))={\mathbb Z}$, static topologically non-trivial
configurations are expected to be present. These configurations are
characterized by an integer number, precisely the Skyrmion number
$N_{sk}$, that in this case is a measure of the wrapping of the
$O(3)$-spin field in the 2-sphere of the compactified Euclidean
space. Even when these configurations exist for the $O(3)$-non-linear
sigma model, it can be proved that they are not energetically favored,
being the ferromagnetic the lowest energy solution. Along the line of
\cite{YON_09}, Zang {\em et. al.}  \cite{ZMH_11} proved that a
non-linear sigma model plus a Dzyaloshinskii-Moriya term and in
presence of a magnetic field, stabilizes a Skyrmions crystal
structure, on some region of parameter space. They obtain a complete
phase diagram from numerical solutions, in which the Skyrmion crystal
phase emerges between the helical and ferromagnetic ones, as the
magnetic field is increased. The values they obtained for the critical
fields are in good agreement with experiments.

In this paper, we present an approximated analytic solution for
the non-linear sigma model with a Dzyaloshinskii-Moriya interaction in the presence of a magnetic field. The expression presented here represents a periodical
arrange of finite volume Skyrmions, and we interpret as the Skyrmion
crystal phase observed in chiral magnets. For every point of
parameter space, an approximated analytic solution can be constructed
in the form of a smoothly pasted internal-external expression. The pasting
condition determines the size of the Skyrmions as a function of the
magnetic field. 

\section{Classical solutions}
\subsection{Non-linear $\Sigma$-model and equations of motion}
The non-linear $\Sigma$-model characterizing the energy of spin variables of a chiral ferromagnet in the square lattice reads
\be
H = \frac12
\int \! d^2x \,
\left(
J\nabla_i n_a \nabla_i n_a+2D\epsilon^{aib}n_a\nabla_i n_b-2B^an_a
\right)\,,
\label{Hamilton}
\ee
where $i=1,2$ and $a=1,2,3$, the vector $n_a$ has unit modulus. Here $J$ is the exchange, $D$ the Dzyaloshinskii-Moriya strength, and $B^a$ the magnetic field, that we assume points in the $3$ direction. 

The unit modulus condition can be solved by writing
\ba
&&n_x=\sin\Theta\cos\Phi\equiv \sin\Theta\, m_x\,,\n
&&n_y=\sin\Theta\sin\Phi\equiv \sin\Theta\, m_y\,,\n
&&n_z=\cos\Theta\,,
\ea
where we have defined the vector $m_i=(\cos\Phi,\sin\Phi)$. Moreover, in what follows we find useful to work with the re-scaled dimensionless spatial coordinates $\underline{x}_i=B_z x_i/2D$ and the re-scaled Hamiltonian $\underline{H}=H/J$. Replacing all this into \eqref{Hamilton} we get the explicit form
\begin{widetext}
\be
\underline{H}= \frac 12
\int \! d^2\underline{x} \,
\left(
\nabla_i \Theta \nabla_i \Theta+\sin^2\!\Theta\, \nabla_i \Phi \nabla_i \Phi
+
p
\left(
\sin\Theta\cos\Theta\,\epsilon^{ij}\nabla_i m_j
+
\epsilon^{ij}m_j\nabla_i \Theta
-
2\cos\Theta
\right)
\right)\,,
\ee
\end{widetext}
where we have defined a parameter $p=4D^2/JB_z$. Notice that, being $p$ the only parameter in the Hamiltonian, any phase transition takes place at a critical value of $p$, let's say $p^{\rm crit}$. Using the explicit form of $p$, we see that the resulting magnetic field is quadratic in the Dzyaloshinskii-Moriya interaction $B^{\rm crit}_z=4D^2/Jp^{\rm crit }$. This implies that in a generic phase diagram the different phases are separated by parabolas in the $B_z$ vs. $D$ plane, the foci of which are determined by the values of $p^{\rm crit}$.

We are interested in configurations that minimize the above energy, the resulting Euler-Lagrange equations read
\ba
&&
\nabla^2\Theta
\!-\!
\sin\Theta
\cos\Theta(\nabla\Phi)^2
\!=\!
p\sin\Theta
\left(
1
\!-\!
\sin\Theta \,m_i\nabla_i\Phi
\right)\,,
\n
&&
\nabla_i\!\left(\sin^2\!\Theta\nabla_i\Phi\right)=
p\sin^2\!\Theta\, m_i\nabla_i\Theta \,.
\label{completas}
\ea

In the forthcoming sections, we will explore the space of solutions of the above equations.
\subsection{Ferromagnetic solution}
Equations \eqref{completas} are solved by the trivial solution 
\be
\Theta_F=0 \,.
\ee
This represents the ferromagnetic configuration, in which all spins are aligned with the magnetic field.

The energy of the ferromagnetic solution read
\be
\underline{H}_F= -p {\cal A}\,,
\ee
where ${\cal A}$ is the dimensionless area of the system.

\subsection{Helix solution}

At zero magnetic field, the above defined system has an helical phase, approximated by the solution
\ba
n_x&=&\cos\beta\,,\n
n_y&=&\sin\beta\,\cos({px}/2)\,, \n
n_z&=&\sin\beta\,\sin({px}/2)\,,
\ea
where $\beta$ is a constant of integration. For finite magnetic field, we can use this expression as an variational approximation of the exact solution. Plugging back into the energy, we get
%
\be
\underline{H}_{H}\!=-\!\sin\!\beta \left(\frac{p^2}8\sin\!\beta 
\!+\!
\frac2L\left(\cos\frac{px_o}2\!-\!\cos\frac{p(x_o\!\!+\!\!L)}2\right)\!\right){\cal A}\,.
\ee
\normalsize
Where $x_o$ is a constant of integration determined by boundary conditions, $L$ is the extension of the sample in the $\underline{x}$ direction, and ${\cal A}$ its dimensionless area. In the large sample size limit, the term proportional to $1/L$ can be discarded.  
Then, the absolute minimum of this expression is reached for $\sin\beta=1$, reading
\be
\underline{H}_{H}=-\frac{p^2}8 {\cal A}\,.
\ee
\subsection{Skyrmion solution}
To obtain the Skyrmion solution, we go into polar coordinates in the plane,  $\underline{x}={r}\cos\varphi$ and $\underline{y}={r}\sin\varphi$, and propose a radial Anzats $\Theta=\Theta({r})$ and $\Phi=\Phi(\varphi)$. (Here and in what follows the reader must keep in mind that the variable $r$ has been re-scaled and it is dimensionless). We get
\ba
&&
r\partial_r\!(r\partial_r\Theta)
\!-\!\!
\sin\!\Theta
\cos\!\Theta(\partial_\varphi\!\Phi)^2
\!=\!
p\sin\!\Theta
\left(
r^2\!\!
\!-\!
\sin\!\Theta \,m_\varphi\partial_\varphi\Phi
\right),
\n
&&
\partial_\varphi^2\Phi=
p\,r^2
m_r\partial_r\Theta \,.
\label{completas-r}
\ea
Here $m_\varphi=r\sin(\Phi-\varphi)$ and $m_r=\cos(\Phi-\varphi)$ are the angular and radial components of the $m_i$ vector, respectively. We see that the second equation in \eqref{completas-r} is solved by $\Phi=\varphi\pm\pi/2$. Replacing into the first, we are left with
\ba
&&
r\partial_r(r\partial_r\Theta)
\!-\!
\sin\Theta
\cos\Theta
\!=\!p\,r
\sin\Theta
\left(
r
\!\mp\!\sin\Theta \right)\,.
\label{sandanga}
\ea

We want to solve the above equation with the boundary condition $\Theta=\pi$ at $r=0$ and $\Theta=0$ at $r=r_b$, where $r=r_b$ identifies the boundary of the Skyrmion. This implies that, as we go from the center to the boundary of the Skyrmion, the vector $n^a$ makes a half-integer number of turns, ending up aligned with the magnetic field.

Solving this equation analytically is not possible. In what follows, we study two different limits that admit analytic solutions, and propose to build a pasted expression made of such analytic solutions in different regions of space, in order to have an approximate solution of the complete system.

\subsubsection{The exchange dominated region and the ring solution}
\begin{figure}[ht]
\center{~\includegraphics[width=.4\textwidth]{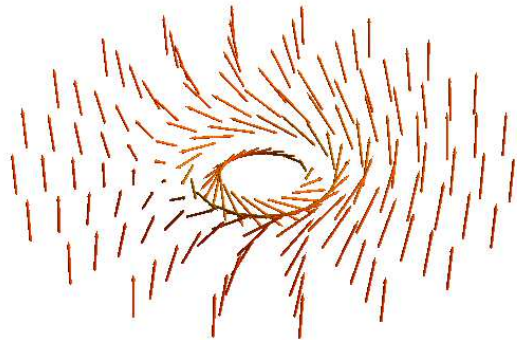}}
\caption{Pictorial representation of the ring solution. The solution has been cutoffed at finite interior radius.}
\label{fig:Ring} 
\end{figure}

If there is a region of space in which the differential equation \eqref{sandanga} is dominated by the contribution coming from the exchange $J$, we can take the limit $p\to 0$ there, and it gets simplified to
\ba
&&
r\partial_r\!\left(r\partial_r\Theta\right)
-
\sin\Theta
\cos\Theta 
=
0\,.
\ea
This equation can be rewritten as
\be
\partial_{\log(r)}^2 \Theta = \sin\Theta\cos\Theta\,.
\ee
Multiplying both sides by $\partial_{\log(r)} \Theta$ we have
\be
\partial_{\log(r)}\left(\partial_{\log(r)}\Theta\right)^2 =  \partial_{\log(r)}(\sin^2\!\Theta)\,,
\ee
from which we get a first integral as
\be
\left(\partial_{\log(r)}\Theta\right)^2 -\sin^2\!\Theta=\frac{1}{k^2}\,,
\label{fi}
\ee      
where $k$ is a constant of integration. This expression allows us to separate variables and perform the remaining integration, as
\be
\log(r_b/r)=k\int^\Theta\!\!\!\!\frac{d\Theta}{\sqrt{1+k^2\sin^2\!\Theta}}=
k F(\Theta|-\!k^2)\,,
\label{ieifk}
\ee
here $r_b$ is a second constant of integration, and $F(\Theta|m)$ is the incomplete elliptic integral of the first kind. In order to have a decreasing $\Theta$ as we approach the boundary where $\Theta=0$, we have to choose $k>0$. Expression \eqref{ieifk} can be inverted, to get
\be
\Theta_R= 
\mbox{am}\!\left(\frac{\log(r_b/r)}{k}|-\!k^2\right)
\,,
\label{Jam}
\ee
where $\mbox{am}(\alpha|m)$ is the inverse of the incomplete elliptic integral of the first kind, also known as the Jacobi amplitude. It can be checked that at $r=r_b$ we have $\Theta_R=0$, thus we consider this solution for $r<r_b$. On the other hand, as we approach the origin the function $\Theta_R$ goes to $-\infty$, implying that the vector $n_a$ does an infinite number of turns. In consequence we must impose an IR cutoff at a finite radius $r=x$. For that reason, we call this solution a ``ring'', and denoted it by a subindex $R$ in equation \eqref{Jam}. A pictoric representation of the solution is shown in Fig.\ref{fig:Ring}.
\vspace{.5cm}

Now we can ask the following question: how good is solution \eqref{Jam} as an approximation of the full solution to equations of motion \eqref{sandanga}, or in other words how far is the right hand side of \eqref{sandanga} from zero? Evaluated on the ring, the absolute value of the right hand side reads
\be
\Delta_{R}\!=\!pr\! \left|
{\rm sn}\!\left(\frac{\log(r_b/r)}{k}{|}\!-\!k^2\!\right)\!\left(\!r\!-\!{\rm sn}\!\left(\frac{\log(r_b/r)}{k}{|}\!-\!k^2\!\right)\right)\right|\,,
\ee
where ${\rm sn}(\alpha,m)$ stands for the Jacobi function defined as the sine of the Jacobi amplitude ${\rm sn}(\alpha,m)=\sin{\rm am}(\alpha,m)$. This is a periodic function of its first argument, that oscillates between $1$ and $-1$. In consequence, the function $\Delta_{R}$ is the sum of two terms that oscillate infinite times as $r\to 0$, their amplitudes being linearly and quadratically decreasing with $r$ respectively. In the forthcoming sections, we compare this error with the one corresponding to a different approximate solution, in order to determine which one is better at a certain radius.
\subsubsection{The Dzyaloshinskii-Moriya dominated region and the vortex solution}
\begin{figure}[ht]
\hspace{5cm}\includegraphics[width=.45\textwidth]{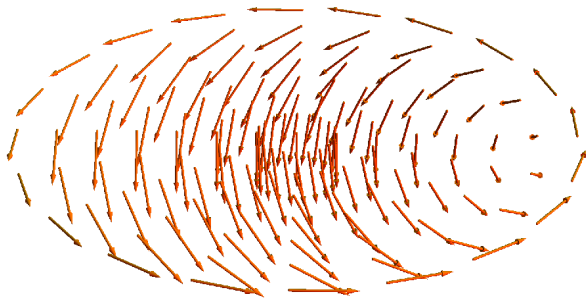}
\caption{Pictorial representation of the vortex solution. Notice that the solution ends at $r=1$ where the spins are completely tangent to the exterior circle.}
\label{fig:Vortex} 
\end{figure}

Conversely, if there is a region of space in which the differential equation \eqref{sandanga} is dominated by the contribution coming from the Dzyaloshinskii-Moriya and Zeeman terms, we can take the limit $p\to \infty$ there, and the equation of motion reads
\ba
&&
r\sin\Theta (r
\mp
\sin\Theta 
)
=0\,.
\ea
This is trivially solved by
\ba
&&
\sin\Theta 
=\pm r\,.
\ea
Since $|\sin\Theta|\leq1$ then $r\leq 1$. Being the polar angle, $\Theta$ is limited to the range $0<\Theta<\pi$ where its sine is positive. This forces us to discard the solution with the $-$ sign. Thus the solution has $\Theta=\pi$ at the origin  and $\Theta=\pi/2$ at a boundary, that sits at $r=1$. We are left with
\ba
&&
\Theta_V
=\pi-\arcsin(r)\,,
\label{vort}
\ea
where $\arcsin(r)$ is assumed to take values in the $\{-\pi/2,\pi/2\}$ range. At the boundary $r=1$, the vector $n_a$ lies completely on the plane, and it is tangent to the boundary circle. That's why we call this solution a ``vortex'', and denoted it with a subindex $V$ in \eqref{vort}. A pictorial representation of the solution is shown in Fig.\ref{fig:Vortex}.

\vspace{.5cm}

As a measure of how far is the above expression to a solution of the full system \eqref{sandanga}, we can use the absolute value of the left hand side of that equation. It reads
\be
\Delta_{V}=
\frac{r^3|2-r^2|}{\sqrt{1-r^2}^3}\,,
\ee
that goes to zero at the origin like $2r^3$.

\subsubsection{Comparison of the errors}

In this section, we identify the regions dominated by the exchange term and by the Dzyaloshinskii-Moriya and Zeeman terms. To do that, we compare the errors $\Delta_{R}$ and $\Delta_{V}$ to have an indication on which one $\Theta_R$ or $\Theta_V$ is the best approximation to the full solution at a given region of the radial variable $r$.  

Given the quotient $\Delta_{R}/\Delta_{V}$, we can say that wherever this quotient is smaller than one, the ring \eqref{Jam} is a better approximation of the full equation \eqref{sandanga} than the vortex. Conversely, wherever the quotient is bigger than one, the best approximation to the solution of the complete system \eqref{sandanga} is given by the vortex \eqref{vort}. Writing the quotient explicitly
\small
\be 
\frac{\Delta_{R}}{\Delta_{V}}
\!\!=\!\!
\frac {p\sqrt{1\!-\!r^2}^3}{|2\!-\!r|r^2} \!\left|
{\rm sn}\!\left(\!\frac{\log(r_b/r)}{k}{|}\!-\!\!k^2\!\!\right)\!\left(\!r\!-\!\!{\rm sn}\!\left(\!\frac{\log(r_b/r)}{k}{|}\!-\!\!k^2\!\!\right)\right)\right|,
\ee 
\normalsize
we see that it it vanishes as $r\to1$, and oscillates infinite times as $r\to0$ with an amplitude that diverges as $r^{-2}$. 

Since as $r\to1$ the quotient goes to zero, we can say that the ring is the best approximation to the solution of the equations of motion (as it better be, since the vortex cannot be extended further than that point). As we move to the interior $r\to0$, the quotient grows and becomes bigger than one at a finite radius, bellow which the vortex is a better approximation. Moving further to the interior, the quotient becomes zero again favoring the ring, and then grows again favoring the vortex, and so on. Notice that, since the amplitude and frequency of the oscillations diverge as we go to the origin, the regions where the quotient is bigger than one (favoring the vortex), become thicker as compared to those where it is smaller than one (favoring the ring).

\begin{figure}[ht]
\includegraphics[width=.23\textwidth]{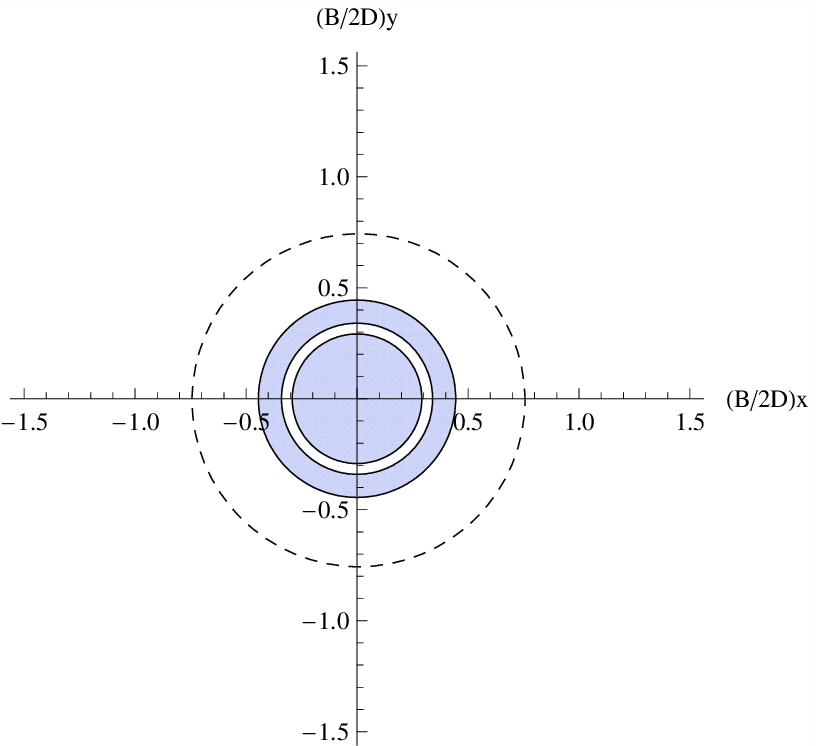}
\includegraphics[width=.23\textwidth]{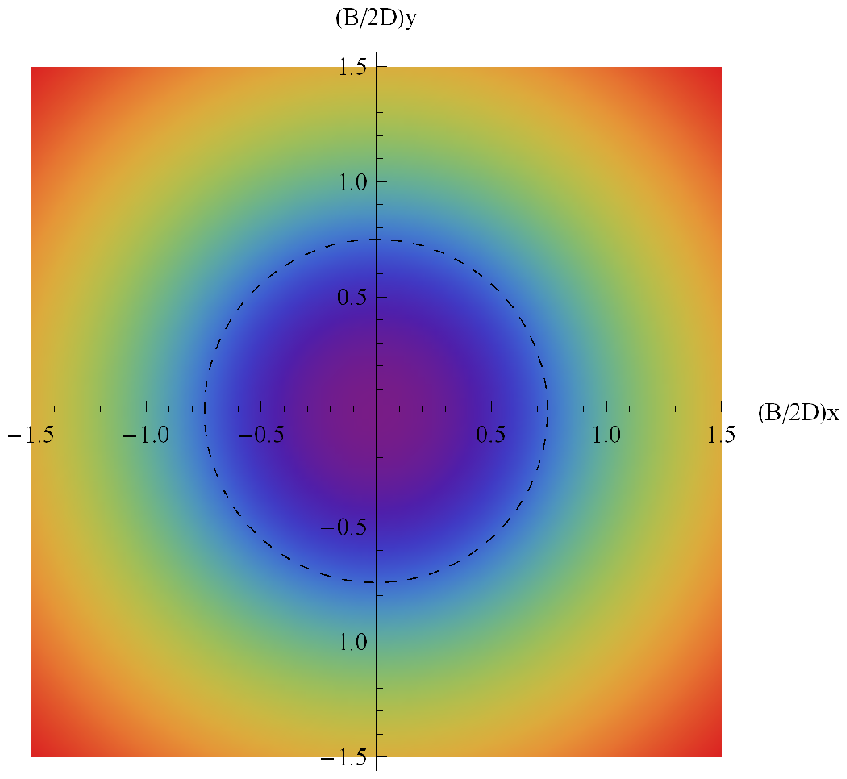}
\caption{Left: representation of the error quotient $\Delta_R/\Delta_V$. In the shaded regions the quotient is bigger than one, favoring the vortex, while in the white regions it is smaller than one, and the ring is better. Right: Skyrmion solution, constructed by pasting an interior vortex and an exterior ring at the dotted line. Notice that, even if in doing so we are extending the vortex outside the shaded region in which it is a better approximation than the ring, this is favored by energy considerations.}
\label{fig:ErrorRegions} 
\end{figure}

The emerging picture is that of a structure of nested rings, as shown in Fig.\ref{fig:ErrorRegions}. The exterior of the solution is well described by the ring, while the interior is well described by the vortex, except in regions whose relative width become very small as we go to the origin. Neglecting those regions, we can try to build a pasted expression, with an exterior made of $\Theta_R$ and an interior made of $\Theta_V$. The radius at which the ring and the vortex are pasted can be considered as a variational parameter in the Hamiltonian. This is what we do in the next subsection.

~ 

\subsubsection{Pasted Skyrmion}

In this section we build a pasted Skyrmion from the building blocks found in the previous sections. We define it as
\be
\Theta_{Sk}=\left\{
\begin{array}{ll}
\Theta_V&\ \ \ \ r\leq x \,,
\\
\Theta_R&\ \ \ \ r>x\,.
\end{array}
\right.
\ee
In order to join smoothly the pieces at $r=x$, we need to impose that the function $\Theta_{Sk}$ and its derivative $\partial_r\Theta_{Sk}$ are continuous at that point. We get the conditions
\ba
\Theta_R&=&\Theta_V\,,
\label{cul}
\\
\partial_r\Theta_R&=&\partial_r\Theta_V\,,
\label{llo}
\ea
or more explicitly
\ba
\mbox{am}\left(\frac{\log(r_b/x)}{k}|-\!k^2\right)&=& \pi-\arcsin\left(x\right)\,,
\label{cu}
\\
\frac1{k
   x}{{\rm dn}\left(\frac{\log
   \left(r_b/x\right)}{k}|-\!k^2\right)}&=&\frac{1}{\sqrt{1-x^2}}\,.
\label{lo}
\ea
Where ${\rm dn}(\alpha,m)$ is the Jacobi function defined as ${\rm dn}(\alpha,m)=\sqrt{1-m\,{\rm sn}^2(\alpha,m)}$. Taking the sine on both sides of the first equation \eqref{cu}, and using this definition on in the second \eqref{lo}, we get
\ba
\mbox{sn}\left(\frac{\log(r_b/x)}{k}|-\!k^2\right)&=&x\,,
\label{an}
\\
1+
k^2 x^2
&=&\frac{{k^2 x^2}}{1-x^2} \,.
\label{no}
\ea
The second equation \eqref{no} can be solved for $k$, obtaining
\be
k(x)= 
\frac{\sqrt{1-x^2}}{x^2}\,,
\ee
while the first  equation \eqref{an} (or equivalently \eqref{cu}) gives a value for $r_b$
\begin{figure}[ht]
\includegraphics[width=8cm]{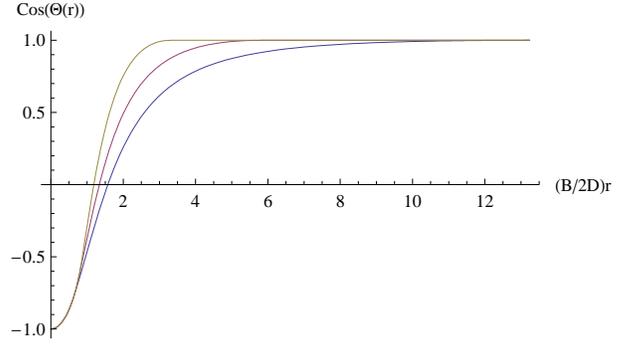}
\caption{Examples of the pasted Skyrmion, for different values of the radius $r_b$, or equivalently of the pasting point $x$. The plots correspond to $r_b=0.6,0.75,0.85$ from top to bottom respectively.}
\label{fig:PastedSolutions} 
\end{figure}
\be
r_b(x)=x\exp\left(
{\frac{\sqrt{1\!-\!x^2}}{x^2} F\!\left(\pi\!-\!\!\arcsin\left(x\right)|
\frac{x^2\!-\!1}{x^4}
\right)}
\right)\,.
\ee
In consequence, the radius of the Skyrmion $r_b$ determines the pasting point $x$ and the integration constant $k$.

Examples of the resulting pasted Skyrmion for different values of the parameters are shown in Fig.\ref{fig:PastedSolutions}.

The energy $\underline{H}_{Sk}$ of the pasted Skyrmion built in this way is given by
\ba
\underline{H}_{Sk}
\!\!&=&\!\! \pi\!\!
\int \! dr \,r\!
\left(
\partial_r \Theta^2\!\!+\!\frac{\sin^2\!\Theta}{r^2}
\!+\!
p\!
\left(\!
\partial_r\Theta 
\!+ \!
\frac{\sin\!2\Theta}{2r}
\!-\!
2
\cos\Theta\!
\right)\!
\right)
\n
&\equiv&\pi r_b^2(x)
\left(
A(x)+p \left(B(x)-2C(x)\right)\right)
\,.
\label{culo2}
\ea
Where  we have defined
\ba
A(x)&=&\frac1{r_b^2(x)} \left(A_1(x)+A_2(x)\right) \,,
\n
B(x)&=&\frac1{r_b^2(x)}\left(B_1(x)+B_2(x)\right) \,,
\n
C(x)&=& \frac1{r_b^2(x)}\left(B_3(x)\!+\!B_4(x)\right)\,,
\label{ogt}
\ea
with
\ba
A_1(x)&=&
\int_0^x \! dr \,r\,
\left((\partial_r \Theta_V)^2+\frac{\sin^2\!\Theta_V}{r^2}\, \right)\,,
\n
A_2(x)&=&
\int_x^{r_b(x)} \! dr \,r\,
\left((\partial_r \Theta_R)^2+\frac{\sin^2\!\Theta_R}{r^2}\, \right)\,,
\label{Alga}
\ea
and
\ba
B_1(x)&=&\int_0^x \! dr \,r\,
\left(
\partial_r\Theta_V
+\frac1{r}\sin\!\Theta_V\cos\!\Theta_V 
\right)\,,
\n
B_2(x)&=&
\int_x^{r_b(x)} \! dr \,r\,
\left(
\partial_r\Theta_R
+\frac1{r}\sin\!\Theta_R\cos\!\Theta_R 
\right)\,,
\n
B_3(x)&=&
\int_0^x \! dr \,r\,\cos\Theta_V\,,
\n
B_4(x)&=&
\int_x^{r_b(x)} \! dr \,r\,\cos\Theta_R\,.
\label{Belga}
\ea
In the above definitions, we have taken into account that the vortex $\Theta_V$ extends from $r=0$ up to $r=x$, while the ring $\Theta_R$ goes from $r=x$ up to $r=r_b(x)$. 

Remarkably, all the above integrals, except $B_4(x)$ can be evaluated explicitly, this is done in Appendix \ref{integrales}. Regarding $B_4(x)$, the integral can be approximated by an analytic expression that is shown in Appendix \ref{putaintegral}.
\subsubsection{Re-scaled pasted Skyrmion}
The expression for the pasted Skyrmion $\Theta_{Sk}$ built in in the previous subsection has an analytic form, approximates the exact solution, and has a variational parameter given by the pasting point $x$. Morevover, its energy can be written as an explicit function of the variational parameter, by means of the integrals \eqref{Alga} and \eqref{Belga} as calculated in the Appendices. In principle, we could now use such pasted Skyrmion to build a skyrmion lattice, write the resulting analytic expression for the lattice energy, and minimize it with respect to $x$. 

As it turns out, the pasted Skyrmion has not enough variational freedom  to obtain a phenomenologically interesting phase diagram. Indeed, even when minimized with respect to the pasting point, the on-shell energy of a Skyrmion lattice is still too large to provide a region of the parameter $p$ in which it is preferred to the ferromagnetic and helix phases. To overcome this problem, in this section, we deform the pasted Skyrmion in order to add an additional parameter while keeping at the same time the aforementioned nice properties. A simple way to do so is to define a re-scaled pasted Skyrmion as
\be
\Theta_{Sk\lambda}(r)=\Theta_{Sk}(\lambda r)\,.
\ee
This re-scaled pasted Skyrmion has its pasting point at $\lambda r=x$ and its boundary at $\lambda r=r_b(x)$. The energy has the same form as in \eqref{culo2}, but with the integrals in \eqref{Alga} and \eqref{Belga} now performed from $0$ to $x/\lambda$ and from $x/\lambda$ to $r_b(x)/\lambda$. Changing variables of integration from $r$ to $\lambda r$, we get
\ba
\underline{H}_{Sk\lambda}
\!&\equiv&\!\pi r_b^2(x)
\left(
A(x)+p \left(\frac1\lambda B(x)-\frac{2}{\lambda^2}C(x)\right)\right)
\,.~~~~~~~
\label{culo3}
\ea
where $A(x)$, $B(x)$ and $C(x)$ are defined as in \eqref{ogt}.

\subsection{Skyrmion lattice solution}

\begin{figure}[ht]
\includegraphics[width=8cm]{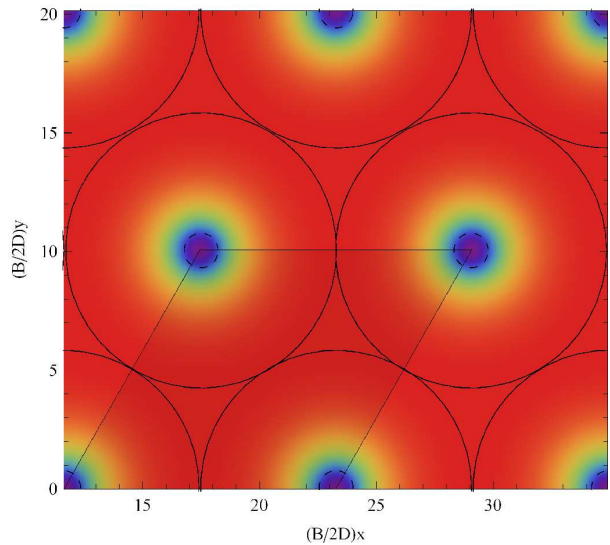}
\caption{Representation of the Skyrmion lattice phase, constructed by compactly packing the Skyrmions and filling the interstitial regions with a ferromagnetic state. Notice that one parallelogram fits a single Skyrmion.}
\label{fig:SkyrmionLattice} 
\end{figure}

Using the re-scaled pasted Skyrmion ($Sk\lambda$) constructed in the previous section, we build the Skyrmion lattice by covering the area of the sample with compactly packed Skyrmions arranged in a triangular lattice. The interstitial space is filled with the ferromagnetic configuration. The resulting configuration can be seen in Fig.\ref{fig:SkyrmionLattice}.

In the Skyrmion lattice, given two triangles forming a parallelogram of side $2r_b(x)/\lambda$ at angle $\pi/3$, it can be seen in the Fig. \ref{fig:SkyrmionLattice} that just one Skyrmion fits in it. Since such a parallelogram has area $2\sqrt 3 r_b^2(x)/\lambda^2$, we have a total of ${\cal A}\lambda^2/2\sqrt 3 r_b^2(x)$ Skyrmions in a sample of area ${\cal A}$. The resulting energy reads
\ba
\!\!\underline{H}_{SkX}\!
&=&\!
\frac{{\cal A}\lambda^2}{2\sqrt3r_b^2(x)}\left(
\underline{H}_{Sk\lambda}
-p(2\sqrt3-\pi)\frac{r_b^2(x)}{\lambda^2}
\right)
\n\!&=&\!
\frac{{\cal A}\pi}{2\sqrt3}\!
\left(
\lambda^2 A(x)+p \left(\lambda B(x)\!-\!2C(x)\!-\!B_5\right)\right),~~
\label{culo}
\ea
where the last term
\ba
B_5= \frac{2\sqrt3}\pi-1 \,,
\ea
is a constant representing the contribution of the interstitial ferromagnetic state.

\begin{figure}[ht]
\setbox1=\hbox{\includegraphics[width=8.5cm]{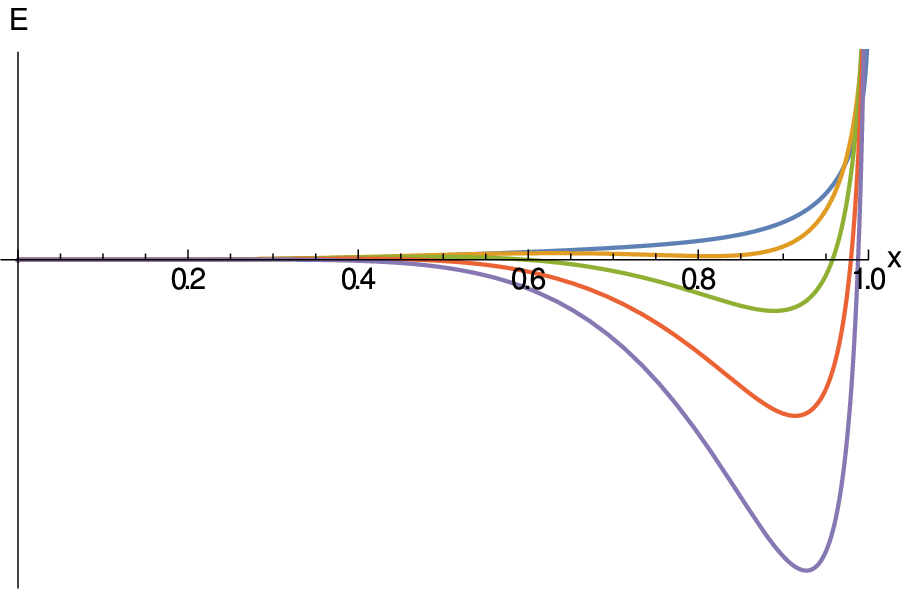}}
\includegraphics[width=8.5cm]{EnergiaVariosp.eps}\llap{\makebox[\wd1][l]{\raisebox{.1cm}{\ \ \ \ \ \includegraphics[height=2.2cm]{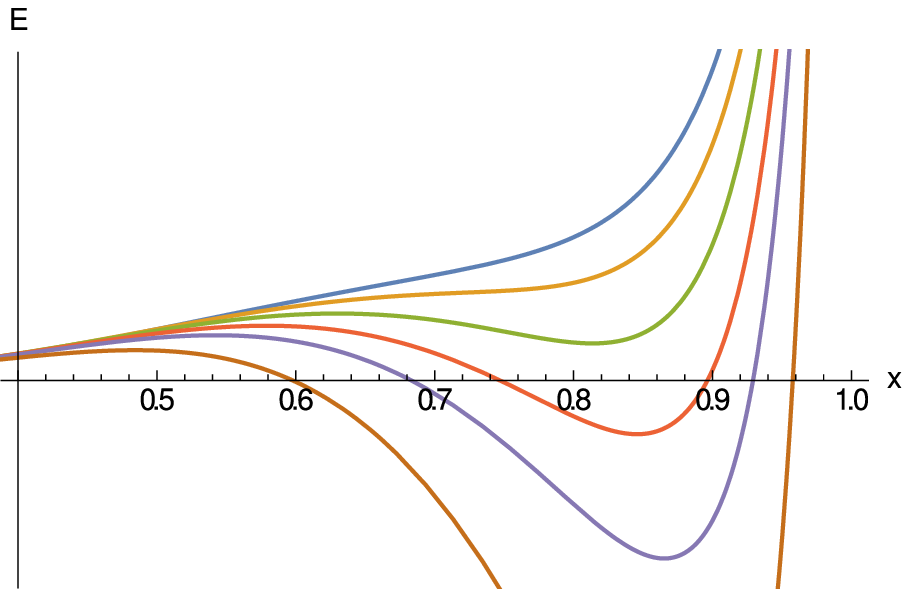}}}}
\caption{Plots of the energy as a function of the pasting radius $x$ for different values of $p$ growing from top to bottom. We see that the absolute minimum moves to the origin at a critical value of $p$. Note: the zero of the energy has been shifted in order to fit all the plots in the same graphic.}
\label{fig:EnergiaVariosp} 
\end{figure}

Expression \eqref{culo} has now to be minimized with respect to the variational parameters $\lambda$ and $x$ in order to obtain the value of the scale $\lambda_m$ and the matching point $x=r_m$ that gives the best approximation. The stationary point condition with respect to $\lambda$ gives the equation 
\be
2\lambda A(x) + p B(x)=0\,,
\ee
which is solved by
\be
\lambda(x) =-\frac{pB(x)}{2A(x)}\,.
\ee
Plugging back into $\underline{H}_{SkX}$ we get the partially on-shell form
\be
\underline{H}_{SkX} =
-\frac{{\cal A}\pi p}{2\sqrt3}\!
\left(
\frac{p}{4}Q(x)  + 2C(x)+B_5\right)\,,
\label{pinQ}
\ee
where we have used the shorthand
\be
Q(x)=\frac{B^2(x)}{A(x)}  \,,
\ee
Fig. \ref{fig:EnergiaVariosp} shows a plot of this energy as a function of $x$ for different values of $p$. As can be seen in the plot, for $p$ large enough the energy has a local minimum at $x=0$, and two stationary points at finite values of $x$. The stationary point at smaller $x$ corresponds to an unstable classical solution of maximum energy (of the kind sometimes called ``thermalon''). On the other hand, the stationary point at larger $x\equiv r_m$ corresponds to the absolute minimum of the energy, that constitutes the true solution. This represents a re-scaled pasted Skyrmion with pasting radius $r_m$. As $p$ decreases, the energy at $x=r_m$ grows until, at a critical value $p_{\rm Collapse}$, it gets greater than the energy at the local minimum at $x=0$. For smaller values of $p$, the absolute minimum is the one sitting at $x=0\equiv r_m$. This represents a ``collapsed vortex'' {\em i.e.} a solution in which the ring extends up to the origin or, in practice, up to a suitable defined UV lattice cutoff $r_m=a$. 

In other words: for $B_z\sim 1/p$ small enough, energetic considerations favor a composite solution, with an exterior ring and an interior vortex pasted at finite radius $r_m$. As the magnetic field grows, a critical value is reached beyond which the vortex collapses, and the favored solution is a pure ring, without any vortex at its interior.  

\begin{figure}[ht]
\includegraphics[width=8.5cm]{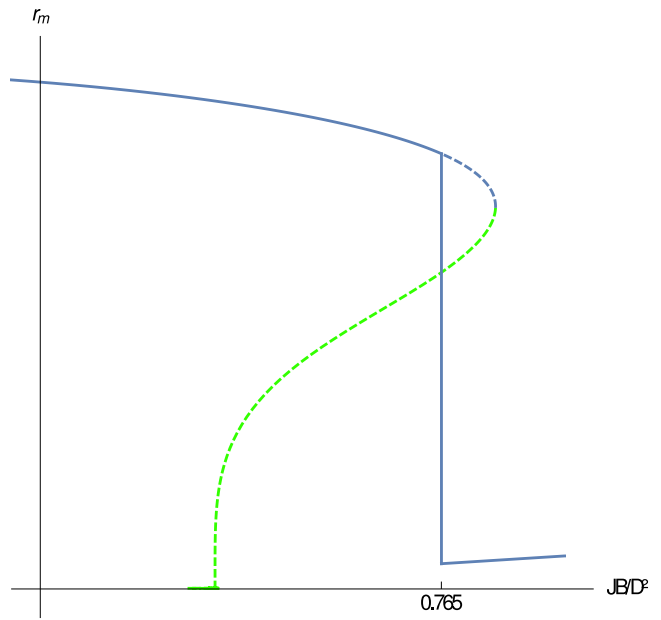}
\caption{Plot of the pasting radius $r_m$ as a function of $4/p=JB_z/D^2$. As we see, there is a critical value at which the vortex collapses leaving us with a pure ring solution.  The continuous blue line is the absolute minimum of the energy, while the dotted blue line is a local minimum that becomes absolute only for $p>p_{\rm Collapse}$. The green dotted line represents a local maximumm the thermalon (see Fig. \ref{fig:EnergiaVariosp}). the non-vanishing value of the pasting radius for the collapsed solution is due to the finite $UV$ lattice cutoff.}
\label{fig:PastingRadius} 
\end{figure}

For $p>p_{\rm Collapse}$, the critical point condition can be written as
\be
\left.\frac{\partial\underline{H}_{SkX}}{\partial x }\right|_{x=r_m}\propto
\frac p4 Q'(r_m)+2 C'(r_m)=0\,,
\ee
where prime $(')$ means derivative with respect to the argument. This equation gives the implicit solution for the matching $r_m$
\be
p=-\frac{8C'(r_m)}{Q'(r_m)}\qquad\qquad p>p_{\rm Collapse} \,.
\label{cho}
\ee
As shown in what follows, this is enough to obtain a phase diagram and we do not need an explicit expression. In other words, we can replace our only parameter $p>p_{\rm Collapse}$ by a new parameter $r_m$ using formula \eqref{cho}.

On the other hand, por $p<p_{\rm Collapse}$, the minimum sits at $r_m=0$. Since the system has a lattice cutoff $a$, we will use in what follows the regularization $r_m=a B_z/2D$. The explicit dependence on the magnetic field appear because of our re-scaling of coordinates. This can be rewritten as 
\be
r_m= \frac\epsilon p\qquad\qquad p<p_{\rm Collapse}\,,
\label{tta}
\ee
where $\epsilon=Ja/2D$ is completely determined by the microscopic parameters.

With this, we can plot the pasting radius $r_m$ as a function of the rescaled magnetic field $(J/D^2)B_z=4/p$. as shown in figure \ref{fig:PastingRadius}.

\section{Phase diagram}

The on-shell energy of the Skyrmion lattice can be written as a function of the pasting point, by replacing the result \eqref{cho} and \eqref{tta} in equation \eqref{pinQ}. We obtain
\be 
\underline{H}_{SkX}\!=\!\!
\left\{
\begin{array}{lll}
\frac{{\cal A}\pi }{\sqrt3}\frac{4C'(r_m)}{Q'(r_m)}\!
\left(
-\frac{2C'(r_m)}{Q'(r_m)}Q(r_m)  \!+\! 2C(r_m)\!+\!B_5\right)\,,
\\~&~&~\\
-\frac{{\cal A}\pi p}{2\sqrt3}\!
\left(
\frac{p}{4}Q(\epsilon/p)  + 2C(\epsilon/p)+B_5\right)
\,.\!
\end{array}
\right.
\ee
where the upper (resp. lower) line corresponds to $p>p_{\rm Collapse}$ (resp. $p<p_{\rm Collapse}$). 

\begin{figure}[ht]
\includegraphics[width=.40\textwidth]{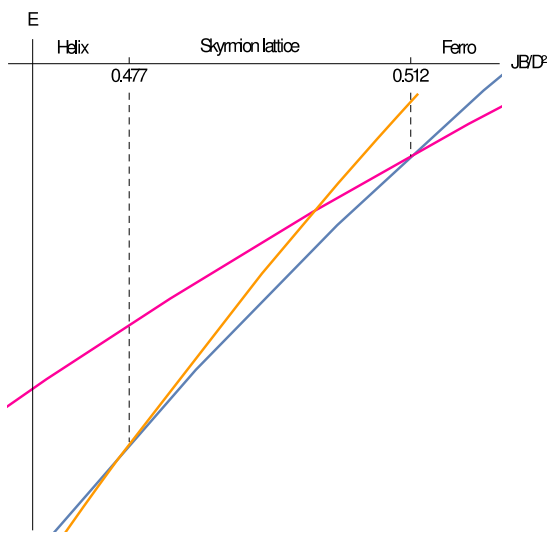}
\caption{Plot of the on shell energy as a function of the magnetic field for the helix, Skyrmion lattice and ferromagnetic solutions. We see that there is an intermediate region in which the Skyrmion lattice phase is favored.}
\label{fig:OnShellEnergy} 
\end{figure}

The energy of the Skyrmion lattice has to be compared with the energy of a ferromagnetic configuration of the same dimensionless area. As it can be seen in the plot Fig.\ref{fig:OnShellEnergy}, they intersect at $4/p=JB_z/D^2=0.512$. For values of $JB_z/D^2$ larger than that, the Skyrmion lattice phase is disfavoured and the ferromagnetic solution dominates. Instead, for smaller values of $JB_z/D^2$, formation of Skyrmions is energetically favoured, and the system will create as many of them as it can accommodate inside its area. This implies that we have a phase transition into a Skyrmion lattice. This phase transition occurs at 
\be
{B_z}^{\rm crit}_F= 0.512 \frac{D^2}J\,.
\label{ferro-skx}
\ee
where we see the critical magnetic field $B_z^{\rm crit}$ with a quadratic dependence in the Dzyaloshinskii-Moriya coupling $D$, as expected on dimensional grounds, with the explicit value of the coefficient.

A similar procedure can be performed to compare the Skyrmion energy with that of the helical phase. Again, we get a critical value of $4/p= JB_z/D^2 =0.477$, which implies
\be
{B_z}^{\rm crit}_H= 0.477 \frac{D^2}J\,.
\label{helix-skx}
\ee
The resulting phase diagram is constructed in the $B$ vs $D$ plane by drawing the parabolas \eqref{ferro-skx} and \eqref{helix-skx}.

\section{Conclusions and outlook}

We have found the zero-temperature phase diagram of a physical system whose relevant dynamics is captured by a nonlinear sigma model with a Dzyaloshinskii-Moriya term and a Zeeman coupling to an external magnetic field. This dynamics corresponds to the continuous limit of an { chiral} ferromagnet on a square lattice.

The diagram presents three phases, helical, Skyrmion lattice, and ferromagnetic, that show up sequentially as the magnetic field is increased.  Dimensional analysis predicts that the critical magnetic field $B_{\rm crit}$ for both phase transitions is quadratic as a function of the Dzyaloshinskii-Moriya coupling $D$ for a fixed value of the exchange strength $J$. We were able to determine the coefficient on such quadratic dependence, our results being consistent with the relevant literature \cite{}.

Our study relies completely on analytic  expressions for the magnetization vector $n^a$. The ferromagnetic phase being trivial, and the variational form of the helical phase being previously known in the literature, the contribution we made in the present work is an approximate analytic expression for the Skyrmion lattice phase. The expression is obtained by compactly packing isolated Skyrmions in a triangular lattice. We obtained an analytic form for the isolated Skyrmion by pasting an interior vortex solution of the Dzyaloshinskii-Moriya Hamiltonian with a Zeeman term, with an exterior obtained as a solution of the pure exchange nonlinear sigma model, and then re-scaling the radial variable to minimize the energy.

This approximation scheme has the advantage of allowing for a perturbative expansion around an analytic form of the background, that could for example be used to study propagation of phonons in the Skyrmion lattice. This is a further development of our approach that we plan to explore in the near future. 

\section{Thanks}

We would like to thank Guillermo Silva, Daniel Cabra, Gerardo Rossini, Fabrizio Canfora, Santiago Osorio, Diego Rosales, Carlos Lamas, and Nicol\'as Kovensky for helpful discussions. { This work is partially supported by grants PIP-2008-0396, PIP-2008-5037 (Conicet, Argentina), PICT-2008-1426  (ANPCyT, Argentina), and PID-2013-X648 (UNLP, Argentina).}

\appendix

\section{Explicit form of the energy integrals}
\label{integrales}
In this Appendix, we will show the explicit computation of the integrals involved in the calculation of the on-shell energy.

The integrals $A_1(x), B_1(x)$ and $B_3(x)$ have to be evaluated on the vortex solution $\Theta_V$ presented on equation \eqref{vort}. We get for $A_1(x)$.
\ba
A_1(x)&=&
\int_0^x \! dr \,r\,
\left((\partial_r \Theta_V)^2+\frac{\sin^2\!\Theta_V}{r^2}\, \right)\,,
\ea
the explicit form
\ba
A_1(x)&=&
\int_0^x \! dr \,r\,
\left(
\frac1{1-r^2}+1 \right)=
\n
&=& \frac12 \left(x^2-\log(1-x^2)\right)\,.
\ea
For $B_1(x)$ instead
\ba
B_1(x)&=&\int_0^x \! dr \,r\,
\left(
\partial_r\Theta_V
+\frac1{r}\sin\!\Theta_V\cos\!\Theta_V 
\right)\,,
\ea
we get 
\ba
B_1(x)&=&\int_0^x \! dr \,r\,
\left(
-\frac1{\sqrt{1-r^2}}-\sqrt{1-r^2}
\right)
\n
&=&
-\frac{1}{3}
\left(
\sqrt{1-x^2} \left(x^2-4\right)+4
\right)\,.
\ea
Finally for $B_3(x)$ the integral reads
\ba
B_3(x)&=&
\int_0^x \! dr \,r\,\cos\Theta_V\,,
\ea
leading to
\ba
B_3(x)&=&
-\int_0^x \! dr \,r\,\sqrt{1-r^2}=
\n
&=&
-\frac{1}{3} \left(1-\sqrt{1-x^2}^3\right)\,.
\ea

~ 

On the other hand, the integrals $A_2(x), B_2(x)$ and $B_4(x)$ have to be evaluated on the ring solution $\Theta_R$ presented on equation \eqref{Jam}. We get for $A_2(x)$.
\ba
A_2(x)&=&
\int_x^{r_b(x)} \! dr \,r\,
\left((\partial_r \Theta_R)^2+\frac{\sin^2\!\Theta_R}{r^2}\, \right)\,,
\ea
changing variable to $\log r$ we have
\ba
A_2(x)\!&=&\!
\int_x^{r_b(x)} \!\! d\log(r) 
\left((\partial_{\log(r)} \Theta_R)^2\!+\sin^2\!\Theta_R\, \right),
\ea
using the first integral \eqref{fi} this can be rewritten as
\ba
A_2(x)&=&
\int_x^{r_b(x)} \! d\log(r) 
\left(\frac 1{k^2(x)}+2\sin^2\!\Theta_R\, \right)\,,
\ea
changing variable to $\alpha=\log(r_b(x)/r)/k(x)$ and replacing the explicit form of the solution \eqref{Jam}, we have
\be
A_2(x)=k(x)
\int^{\bar{\alpha}(x)}_{0} \! d\alpha 
\left(\frac 1{k^2(x)}+2\,{\rm sn}^2\!\left(\alpha|\!-\!k^2(x)\right)\right)\,,
\label{cuuu}
\ee
where we have defined
\ba
\bar \alpha(x)&=&\frac1{k(x)} \log(r_b(x)/x)=
\n
&=&F\!\left(\pi\!-\!\!\arcsin\left(x\right)|
\frac{x^2\!-\!1}{x^4}
\right)\,.
\ea
The first term in \eqref{cuuu} can be integrated trivially, while the integral of the second can be found on tables or performed with {\it Mathematica}. The result is
%
\ba
A_2(x)&=&\frac {x^2}{\sqrt{1\!-\!x^2}}
\left(2
E\left(
\pi\!-\!\!\arcsin\left(x\right)|
\frac{x^2\!-\!1}{x^4}\right)-
\right.\n&&\left.
~~~~~~~~~
-
F\!\left(\pi\!-\!\!\arcsin\left(x\right)|
\frac{x^2\!-\!1}{x^4}
\right)\right).
\ea
%
Now  we turn into $B_2(x)$. It reads
\ba
B_2(x)&=&
\int_x^{r_b(x)} \! dr \,r\,
\left(
\partial_r\Theta_R
+\frac1{r}\sin\!\Theta_R\cos\!\Theta_R 
\right),\ \quad
\ea	
after the same change of variables, we get
%
\ba
B_2(x)\!&=&\!\!
r_b(x)\!\!\int^{\bar\alpha(x)}_{0} \!\!\!\!\!\!\! d\alpha \,e^{-\!k(x)\alpha}
\left(\!-\!
\partial_{\alpha}\!\Theta_R
\!+\!
k(x)\sin\!\Theta_R\cos\!\Theta_R 
\right)\,,\n
\ea	
%
Using the explicit form of the ring solution \eqref{Jam} we can write  it as
%
\ba
B_2(x)&=&
r_b(x)\!\!\int^{\bar\alpha(x)}_{0} \!\!\!\! d\alpha \,e^{-\!k(x)\alpha}
\left(\!-
{\rm dn}\left(\alpha|\!-\!k^2(x)\right)+
\right.\n&&\left.+
k(x){\rm sn}\left(\alpha|\!-\!k^2(x)\right){\rm cn}\left(\alpha|\!-\!k^2(x)\right)
\right)=
\n
&=&
r_b(x)\!\!\int^{\bar\alpha(x)}_{0} \!\!\!\! d\alpha \,e^{-\!k(x)\alpha}
\left(\!-
{\rm dn}\left(\alpha|\!-\!k^2(x)\right)
+\phantom{\frac12}
\right.\n&&\left.+
\frac 1{k(x)}
\partial_\alpha{\rm dn}\left(\alpha|\!-\!k^2(x)\right)
\right)=
\n
&=&
\frac {r_b(x)}{k(x)}
\left(
e^{-\!k(x)\bar\alpha(x)}
{\rm dn}\left(\bar\alpha(x)|\!-\!k^2(x)\right)-1
\right),~~
\ea	
%
where in the second line we used the property $\partial_\alpha{\rm dn}\left(\alpha|m\right)=-m\,{\rm cn}\left(\alpha|m\right){\rm sn}\left(\alpha|m\right)$. This expression gets much simpler when we use the fact that ${\rm dn}\left(\alpha|m\right)=\sqrt{1-m\,{\rm sn}^2\left(\alpha|m\right)}$ and the explicit form of $\bar \alpha(x)$. It reads
\ba
B_2(x)
&=&
\frac {x^2}{\sqrt{1-x^2}}
\left(1
-xe^{\frac{\sqrt{1\!-\!x^2}}{x^2}F\left(\pi\!-\!\arcsin(x)|\frac{x^2\!-\!1}{x^4}\right)}
\right)\!\!.\qquad\ 
\ea	

Finally, we have the integral $B_4(x)$, that reads
\ba
B_4(x)&=&
\int_x^{r_b(x)} \! dr \,r\,\cos\Theta_R\,,
\ea
written in terms of the variable $\alpha$, we get
\ba
B_4(x)&=&k(x)
\int^{\bar \alpha(x)}_{0} \! \! \! \! d\alpha \,e^{-2k(x)\alpha}\,{\rm cn}
\left(
\alpha|\!-\!k^2(x)
\right)\,,\quad
\ea
where ${\rm cn}(\alpha|m)$ is the Jacobi function defined as the cosine of the Jacobi amplitude ${\rm cn}(\alpha|m)=\cos {\rm am}(\alpha|m)$. 
We were unable to find an explicit form of this integral. In the next Appendix we develop and approximation scheme for it.

\section{Approximation of the ring magnetic energy}
\label{putaintegral}
The remaining integral represents the magnetic energy, and it reads
\ba
B_4(x)&=&k(x)
\int^{\bar\alpha(x)}_{0} \! \! \! \! d\alpha \,e^{-2k(x)\alpha}\,{\rm cn}
\left(
\alpha|\!-\!k^2(x)
\right)\,.\quad
\label{cullll}
\ea
In what follows we will try to argue that the ${\rm cn}(\alpha|-\!k^2)$ can be replaced by $\cos(\alpha)$ in the integrand.

First, when $x$ is close to $1$ we can use the expansion of the Jacobi amplitude as in powers of its modulus, as
\be
{\rm am}\left(\alpha|m\right)\simeq 
\alpha +\frac{1}{8} m (\sin (2 \alpha )-2 \alpha )\,.
\ee
The relative weight of the first order correction with respect to the zeroth order is $|m|{|(\sin (2 \alpha )-2 \alpha)|}/{8\alpha} < 0.31|m|$. Then, the order $m$ correction will be negligible as long as $0.31|m|\ll1$. In our case $m=-k^2(x)$ which implies $k\ll 1.82$ or $0.65\ll x<1$. In this range, we can safely replace ${\rm cn}\left(\alpha|\!-\!k^2(x)\right)=\cos{\rm am}\left(\alpha|\!-\!k^2(x)\right)$ by $\cos(\alpha)$.

In the complementary region $x<0.65$, we must try a different justification. Due to the exponential factor, the integrand vanishes for $\alpha$ large enough. More precisely, the exponential is smaller than $\varepsilon$ for $\alpha>|\log\varepsilon|/2k(x)$. Then, by choosing $\varepsilon$ small enough, we can replace the upper integration limit by $|\log\varepsilon|/2k(x)$ and discard the contribution of the rest of the integration region. Now, the function ${\rm cn}\left(\alpha|\!-\!k^2(x)\right)$ is periodic with period $K(-\!k^2(x))$, where $K(m)$ is the complete elliptic integral of the first kind. We chose $\varepsilon$ such that this function is approximately constant in the region of integration, namely $K(-\!k^2(x))\gg |\log\varepsilon|/2k(x)$. Since $K(-\!k^2(x))<2\pi$, the function $\cos(\alpha)$ is also approximately constant in that region. We can then interchange ${\rm cn}\left(\alpha|\!-\!k^2(x)\right)$ by $\cos(\alpha)$ safely. The condition $K(-\!k^2(x))\gg |\log\varepsilon|/2k(x)$ implies  $K(-\!k^2(x))2k(x)\gg |\log\varepsilon|$, which in the region $x<0.65$ ({\em i.e.} $k>1.82$) requires $5.52\gg |\log\varepsilon|$ or $\varepsilon\gg 0.0041$.

Regarding the errors made with the approximation, the contribution of the discarded region is of order $\varepsilon |\bar\alpha(x)-|\log\varepsilon|/2k(x)|< 0.013$. Moreover the error made by replacing ${\rm cn}(\alpha|m)$ by $\cos(\alpha)$ is of order $0.025$. In conclusion, up to the second decimal digit, we can replace the integral \eqref{cullll} by
\ba
B_4(x)&=&k(x)
\int^{\bar\alpha(x)}_{0} \! \! \! \! d\alpha \,e^{-2k(x)\alpha}\,\cos(
\alpha)
\n
&=&
\frac{e^{-2 \bar \alpha(x)  k(x)} (\sin (\bar \alpha(x) )-2 k \cos (\bar \alpha(x) ))+2k(x)}{4 k^2(x)+1}.
\n
\label{cullllee}
\ea

\bibliography{PerseveraYTriufaras.bib}
%


\end{document}